\documentclass{PoS}

\title{Two-flavor lattice QCD study of $\Lambda$(1405)}

\ShortTitle{Two-flavor lattice QCD study of $\Lambda$(1405)}

\author{\speaker{Toru T. Takahashi}\\
        Yukawa Institute for Theoretical Physics, Kyoto University,
        Sakyo, Kyoto 606-8502, Japan\\
        E-mail: \email{ttoru@yukawa.kyoto-u.ac.jp}}

\author{Makoto Oka\\
        Department of Physics, H-27, Tokyo Institute of Technology, 
        Meguro, Tokyo 152-8551 Japan\\
        E-mail: \email{oka@th.phys.titech.ac.jp}}

\abstract{
Low-lying $\Lambda$ baryons with spin 1/2 are 
analyzed in two-flavor unquenched lattice QCD.
We construct $2 \times 2$ cross correlators 
from flavor SU(3) ``octet'' and ``singlet'' baryon operators,
and diagonalize them so that we can
extract two low-lying states for each parity.
The two-flavor CP-PACS gauge configurations are employed, 
which are generated in the renormalization-group improved gauge action
and the ${\mathcal O}(a)$-improved quark action.
Simulation are performed at three different $\beta$'s, 
$\beta = 1.80$, 1.95 and 2.10,
whose corresponding lattice spacings are $a = 0.2150$, 0.1555 and 0.1076 fm. 
For each cutoff, we adopt four different hopping parameters,
($\kappa_{\rm val}, \kappa_{\rm sea}$).
The corresponding pion masses range about from 500 MeV to 1.1 GeV.
Results indicate that 
there are two negative-parity $\Lambda$ states nearly degenerate 
at around 1.6 GeV, while no state as low as $\Lambda (1405)$ is observed.
By decomposing the flavor components of each state, 
we find that the lowest (1st-excited) negative-parity state
is dominated by flavor-singlet (flavor-octet) component.
}

\FullConference{The XXVII International Symposium on Lattice Field Theory - LAT2009\\
		 July 26-31 2009\\
		 Peking University, Beijing, China}

\begin{document}

\section{Introduction}

$\Lambda (1405)$ has been attracting much interest from several view points. 
$\Lambda (1405)$ is the lightest negative-parity baryon, 
even though it has one valence strange quark in it.
Among the $J^P = 1/2^-$ baryons, $\Lambda (1405)$ is 
isolated from the others, much lighter than
the non-strange counterpart $N(1535)$.
It has no spin-orbit partner in the vicinity, as the lowest spin $3/2^-$
state is $\Lambda (1520)$.
Furthermore, the structure of $\Lambda (1405)$ remains mysterious. 
On one hand, $\Lambda (1405)$ is interpreted as a flavor-SU(3)-singlet 
three-quark state in conventional quark models.
On the other hand, 
$\Lambda (1405)$ could be interpreted as an antikaon-nucleon 
$\bar KN$ molecular bound state (B.E. $\sim$ 30 MeV). 
The binding energy of $\bar KN$ implies a
strong attractive force between $\bar K$ and $N$
~\cite{Sakurai:1960ju,Dalitz:1967fp}, 
which may cause a new type of dense hadronic
matter, kaonic nuclei or kaonic nuclear matter
~\cite{Akaishi:2002bg,Yamazaki:2002uh,Akaishi:2005sn}.
We also expect that  such $\bar KN$ bound states with large
binding energies can be regarded as compact 5-quark states.
The 5-quark picture of $\Lambda(1405)$ has advantages
that all five quarks can be placed in the lowest-lying $L=0$ state
to form a negative-parity baryon, and also 
that it requires no spin-orbit partner of $\Lambda (1405)$.

The property of $\Lambda (1405)$ 
can therefore be an important clue to new paradigm in hadron physics. 
We here study properties of $\Lambda(1405)$ using the lattice QCD formulation.
Though several lattice QCD studies
on $\Lambda (1405)$ have been performed so far
~\cite{Melnitchouk:2002eg,Nemoto:2003ft,Burch:2006cc,Ishii:2007ym},
most of them are based on quenched QCD and 
few lattice QCD studies succeeded in reproducing the mass of $\Lambda (1405)$.
Moreover, in many cases, individual analyses with ``singlet'' and ``octet''
operators were performed,
where possible flavor mixings were not taken into account.
Little has been discussed on the lattice about the possible mixing
of flavor-SU(3)-octet and -singlet components induced by
the symmetry breaking.
It is then an intriguing issue to clarify the flavor structures
in excited-state hadron resonances.

Several possible reasons 
for the failure of reproducing $\Lambda (1405)$ in lattice QCD
were suggested through these studies, such as
missing meson-baryon components due to quenching,
exotic (non-3-quark type) structure of $\Lambda (1405)$,
or insufficiency of the lattice volume in the simulations.
Resolving such difficulties requires 
unquenched lattice QCD calculation on a larger lattice volume
with varieties of interpolating operators.
Then, evaluating contamination by scattering states
induced purely by sea quarks would be newly needed.
In this report, we aim at clarifying the properties of $\Lambda (1405)$
with two-flavor full lattice QCD, adopting the ``octet'' and ``singlet''
baryon operators to construct correlation matrices,
which enables us to extract the low-lying spectrum
as well as the mixing between octet and singlet
components in $\Lambda (1405)$.

\section{Lattice QCD setups}

\subsection{simulation conditions}

We adopt the renormalization-group improved gauge action
and the ${\mathcal O}(a)$-improved quark action.
We adopt three different $\beta$'s,
$\beta = 1.80$, 1.95 and 2.10,
and corresponding lattice spacings are $a = 0.2150$, 0.1555 and 0.1076
fm~\cite{AliKhan:2001tx}.
We employ four different hopping parameters 
($\kappa_{\rm val}, \kappa_{\rm sea}$)
for each cutoff.
Corresponding pion masses range approximately from 500MeV to 1.1 GeV
at each $\beta$.
The details are found in Ref.~\cite{Takahashi:2009bu}

\subsection{baryonic operators}

In order to extract the low-lying states in $S=-1$ and isosinglet channel,
we construct $2\times 2$ cross correlators
from the ``singlet'' and ``octet'' operators,
\begin{eqnarray}
{\eta}_{\bf 1}(x)
\equiv
\frac{1}{\sqrt{3}}
\epsilon^{abc}
\left(
u^a(x)[d^{T b}(x) C \gamma_5 s^c(x)] 
+
d^a(x)[s^{T b}(x) C \gamma_5 u^c(x)]
+
s^a(x)[u^{T b}(x) C \gamma_5 d^c(x)] 
\right) \nonumber
\label{defeta1}
\end{eqnarray}
\begin{eqnarray}
{\eta}_{\bf 8}(x)
\equiv
\frac{1}{\sqrt{6}}
\epsilon^{abc}
\left(
u^a(x)[d^{T b}(x) C \gamma_5 s^c(x)]
+
d^a(x)[s^{T b}(x) C \gamma_5 u^c(x)]
-2
s^a(x)[u^{T b}(x) C \gamma_5 d^c(x)]
\right) \nonumber
\label{defeta2}
\end{eqnarray}
It is easy to check that ${\eta}_{\bf 1}(x)$
(${\eta}_{\bf 8}(x)$) belongs to the singlet (octet) 
irreducible representation of the flavor SU(3).
We adopt point-type operators for the sink, $\eta(x)$,
and extended operators, which are smeared
in a gauge-invariant manner, for the source, $\bar\eta(y=0)$.
Smearing parameters are chosen so that 
root-mean-square radius is approximately 0.5 fm.

\section{Lattice QCD Results}

\subsection{hadronic masses}

\begin{figure}[hbt]
\begin{center}
\includegraphics[scale=0.28]{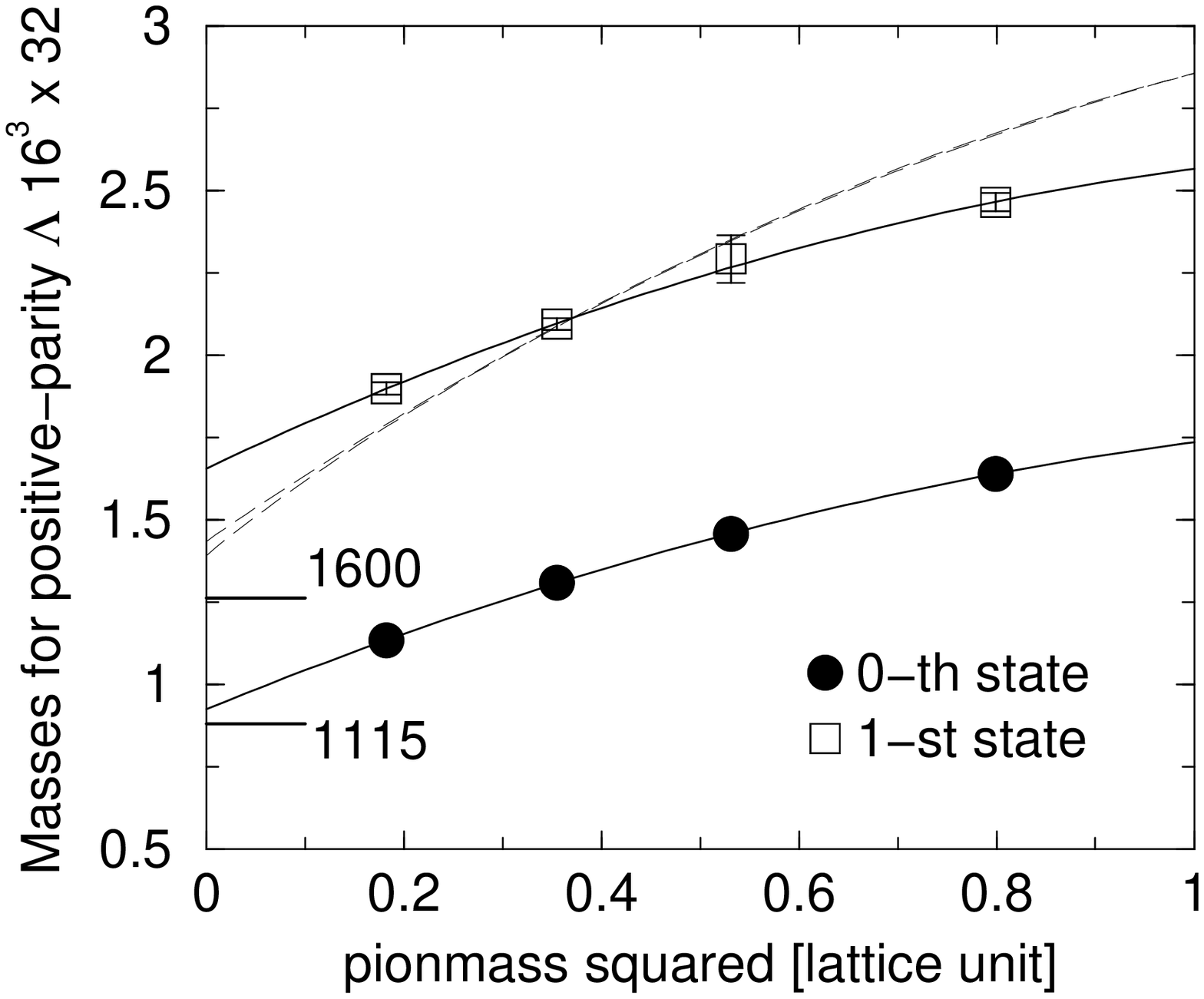}
\includegraphics[scale=0.28]{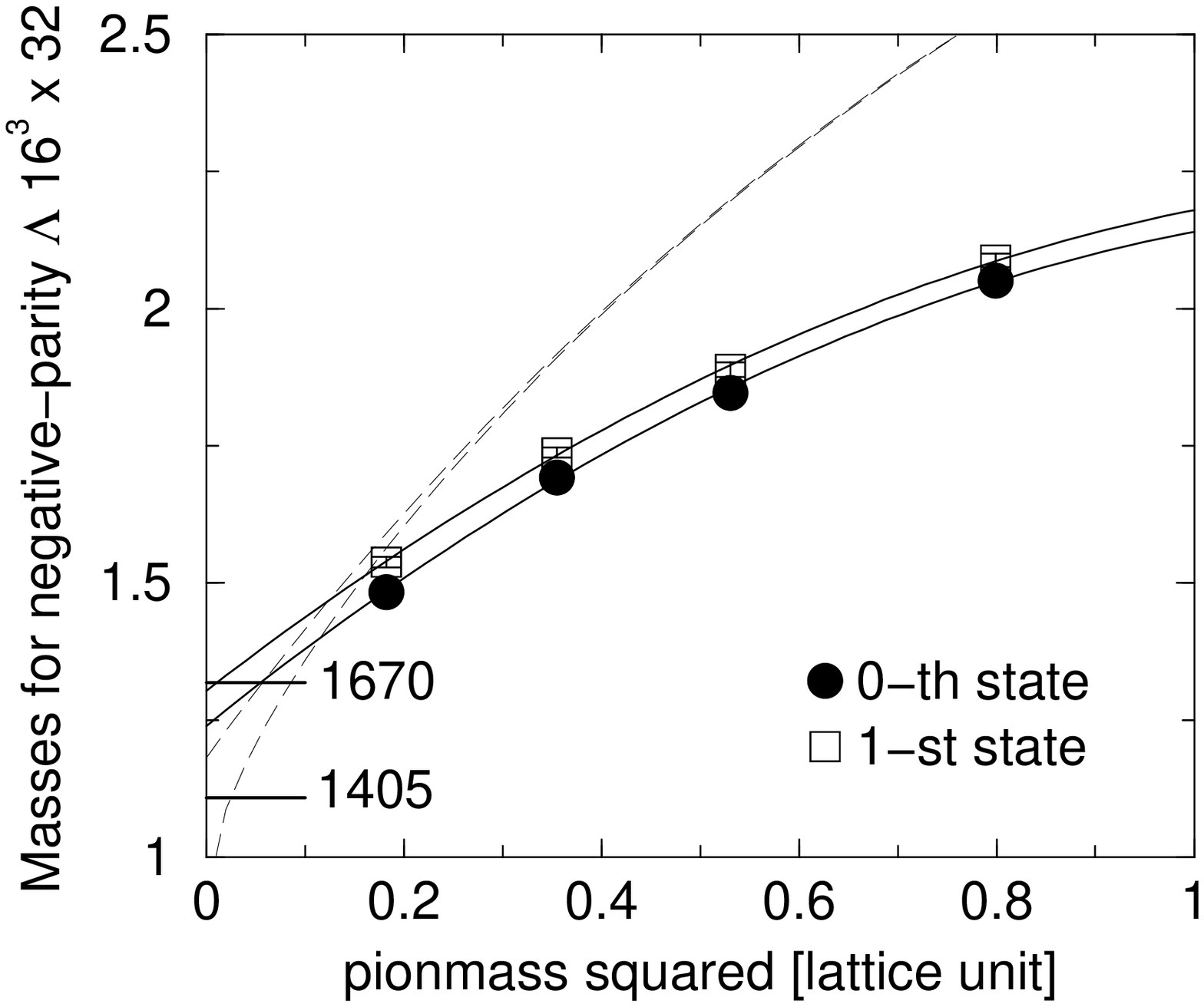}
\includegraphics[scale=0.30]{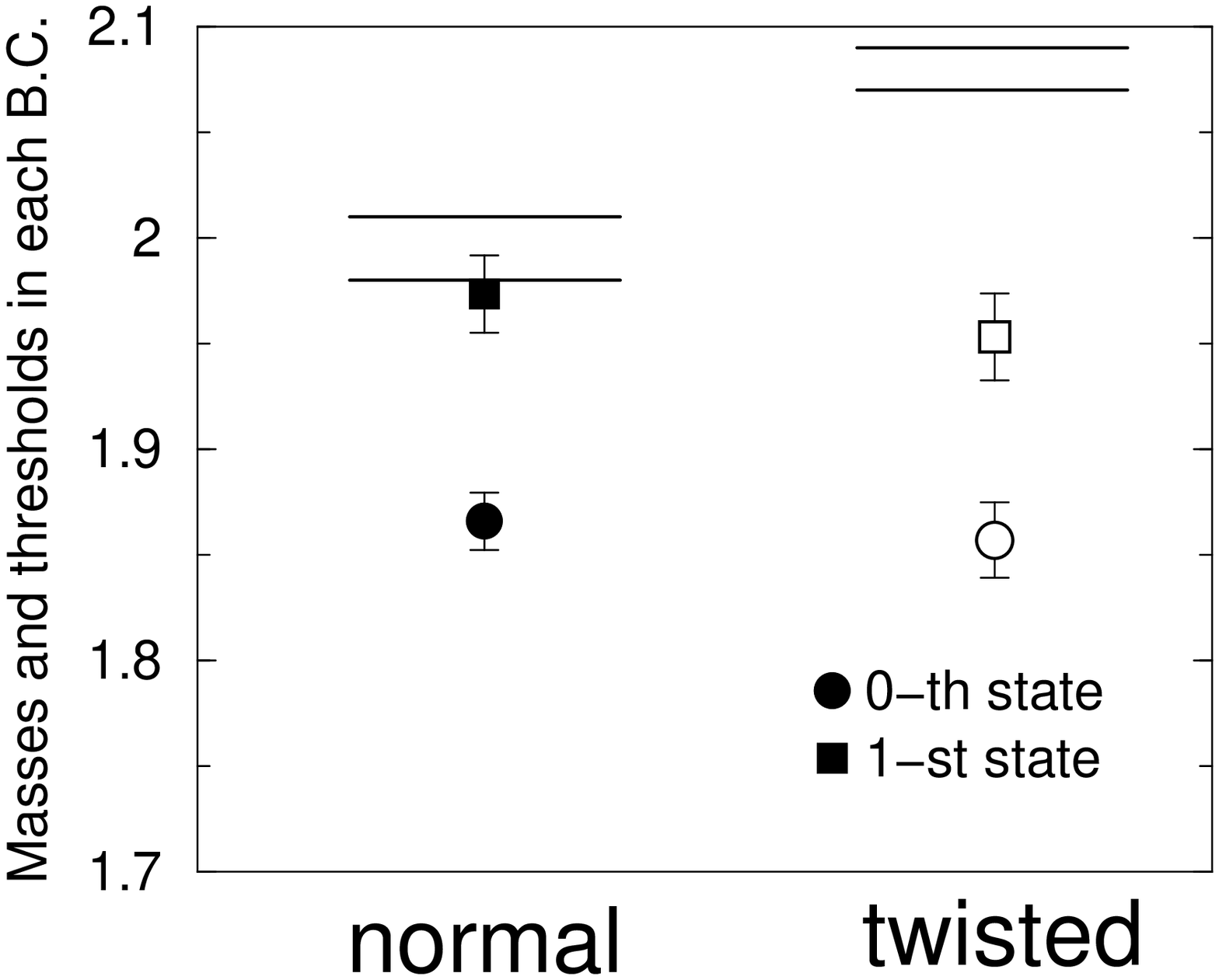}
\end{center}
\caption{\label{posmass}
Masses of the lowest two $\Lambda$ states plotted as functions
of the squared pion mass.
The filled circles (open squares) denote
the masses of lowest (1st-excited) state.
Two solid curves represent quadratic functions
used in the chiral extrapolation.
({\bf left panel}):
Two dashed lines indicate the $\pi\Sigma$ and the $\bar K N$ thresholds
in the presence of the relative momentum $p=\frac{2\pi}{L}$.
Two solid lines on the vertical axes show
the experimentally observed masses of $\Lambda(1115)$ and $\Lambda(1600)$.
({\bf middle panel}):
Two dashed lines indicate the $\pi\Sigma$ and the $\bar K N$ thresholds
with the relative momentum $p=0$.
Two solid lines on the vertical axes show
the experimentally observed masses of $\Lambda(1405)$ and $\Lambda(1670)$.
({\bf right panel}):
Lowest two eigen-energies in the $(J^P,S) = (1/2^-,-1)$ channel
under the normal (periodic) and the twisted boundary conditions.
The circles (squares) are for the ground (1st-excited) states.
The lower and upper solid lines respectively represent
the threshold energies of $\Sigma\pi$ and $N\bar K$ states
evaluated with normal/twisted boundary condition.
}
\end{figure}

Fig.~\ref{posmass} shows
the eigen-energies in the positive- and negative-parity channels,
plotted as functions of the squared pion mass $m_\pi^2$.
Filled circles denote the energies of the ground states,
and open squares those of 1st-excited states.
The solid curves represent quadratic fits 
as a function of squared pion mass $m_\pi^2$.
Two solid lines at the vertical axes
indicate the masses of $\Lambda(1115)$ and $\Lambda(1600)$
in the left panel (positive-parity states),
and $\Lambda(1405)$, $\Lambda(1670)$ and $\Lambda(1800)$
in the right panel (negative-parity states).

The obtained masses of  the positive-parity ground state
agree very well with the mass of the ground-state $\Lambda(1115)$.
On the other hand, the 1st-excited state in this channel
lies much higher than $\Lambda(1600)$,
which is the 1st excited state experimentally observed so far.
The same tendency was reported in Ref.~\cite{Burch:2006cc}, and 
the situation is similar to the case of  
the Roper resonance, 
which is the non-strange SU(3) partner of $\Lambda (1600)$~\cite{Sasaki:2001nf}.

In the negative-parity channel (The right panel in Fig.\ref{posmass}),
the ground- and the 1st-excited states always have close energies
at all the $\kappa$'s.
The eigen-energies have similar quark-mass dependences, and
the mass splittings are almost quark mass independent.
The chirally extrapolated values both lie around 
the mass of $\Lambda(1670)$ rather than $\Lambda(1405)$.
Similarly to the previous studies,
the mass of $\Lambda(1405)$ is not reproduced in our calculation.
While in quenched simulations in Refs.\cite{Melnitchouk:2002eg,Nemoto:2003ft}
such failure was regarded as an evidence
of possible meson-baryon molecule components in $\Lambda(1405)$,
our present {\it unquenched}
simulation contains effects of dynamical quarks
and thus should incorporate meson-baryon molecular states.

We here discuss possible contaminations from scattering states.
Since our calculations contain dynamical u and d quarks,
scattering states could come into spectra.
($N\bar K$ and $\Sigma\pi$ thresholds
are drawn in Fig.~\ref{posmass} as dashed lines.)
The energy of the 1st-excited state 
at the lightest u- and d-quark masses
could be contaminated by such scattering states.
There are two meson-baryon channels relevant in the present calculation,
$\pi\Sigma$ and $\bar K N$.
In terms of the valence quark contents, we have 5 different thresholds
(2 for $\bar K N$, 3 for $\pi\Sigma$).
In this report, 
in order to distinguish resonance states from all these scattering states,
we impose the following boundary condition on the quark fields,
\begin{equation}
\psi(x+L)
=
e^
{\frac23\pi i}
\psi(x).
\end{equation}
Under such boundary condition for quark fields,
a hadronic state $\phi_{3k+n}(x)$ which contains $3k+n$ valence quarks obeys
\begin{equation}
\phi_{3k+n}(x+L)
=
e^
{\frac23n\pi i}
\phi_{3k+n}(x),
\end{equation}
and can have spatial momenta,
$
p_{\rm lat}
=
\frac{2\pi}{L}m+\frac{2n\pi}{3L}
\ \ 
(m \in {\rm Z})
$.

As a result, only states which consist of 3k valence quarks
can be zero-momentum states.
Since other quark combinations 
inevitably have non-vanishing spatial momenta, their energies are raised up.
As long as we employ three-quark operators for baryon creation/annihilation,
hadronic states appearing in scattering states should contain
one or two valence quark(s), 
since sea-quark pairs themselves cannot carry flavors.

We plot in Fig.~\ref{posmass} the eigen-energies under 
the periodic and the twisted boundary conditions.
The open and filled circles (squares) are for the ground (1st-excited) states.
The lower and upper solid lines respectively represent
the threshold energies of $\Sigma\pi$ and $N\bar K$ states
with normal/twisted boundary condition.
The threshold energies are raised up
in the case of twisted boundary condition,
whereas the lowest two eigen-energies remain unchanged.
Thus we conclude that the observed states are insensitive to boundary conditions,
and contaminations from meson-baryon scattering states are small,
which would be the same reason as the absence of string breaking
in heavy-quark potentials from Wilson loops.

\subsection{Flavor structures}

The chiral unitary approach~\cite{Jido:2003cb} has suggested
that $\Lambda (1405)$ is not a single pole
but a superposition of two independent resonance poles.
The structure of $\Lambda (1405)$
is now attracting much interest,
and desired to be clarified in a model independent manner.
We investigate the flavor structures
of the ground and the 1st-excited states
obtained from the cross correlators of two operators.

In order to clarify the flavor structures of the low-lying states,
we evaluate the overlaps of the obtained states with the source and
sink operators.
We evaluate $g_0$ and $g_1$ defined as
\begin{eqnarray}
g_0^-
&\equiv&
\langle {\rm 0th} | \eta_{\bf 8} | {\rm vac}\rangle
/
\langle {\rm 0th} | \eta_{\bf 1} | {\rm vac}\rangle
\label{gminus0}
\\
g_1^-
&\equiv&
\langle {\rm 1th} | \eta_{\bf 1} | {\rm vac}\rangle
/
\langle {\rm 1th} | \eta_{\bf 8} | {\rm vac}\rangle
\label{gminus1}
\end{eqnarray}
Both $g_0^-$ and $g_1^-$ vanish when the SU(3)$_F$ symmetry is exact,
showing that the ground (1st excited) state is purely flavor singlet (octet) 
in the limit.
For the positive-parity states, we similarly define
\begin{eqnarray}
g_0^+
&\equiv&
\langle {\rm 0th} | \eta_{\bf 1} | {\rm vac}\rangle
/
\langle {\rm 0th} | \eta_{\bf 8} | {\rm vac}\rangle
\label{gplus0}
\\
g_1^+
&\equiv&
\langle {\rm 1th} | \eta_{\bf 8} | {\rm vac}\rangle
/
\langle {\rm 1th} | \eta_{\bf 1} | {\rm vac}\rangle
\label{gplus1}
\end{eqnarray}
In this case, we exchange the denominator and the numerator
since the ground state is flavor octet and
the 1st excited state is flavor singlet in the SU(3)$_F$ limit.

\begin{figure}[hbt]
\begin{center}
\includegraphics[scale=0.35]{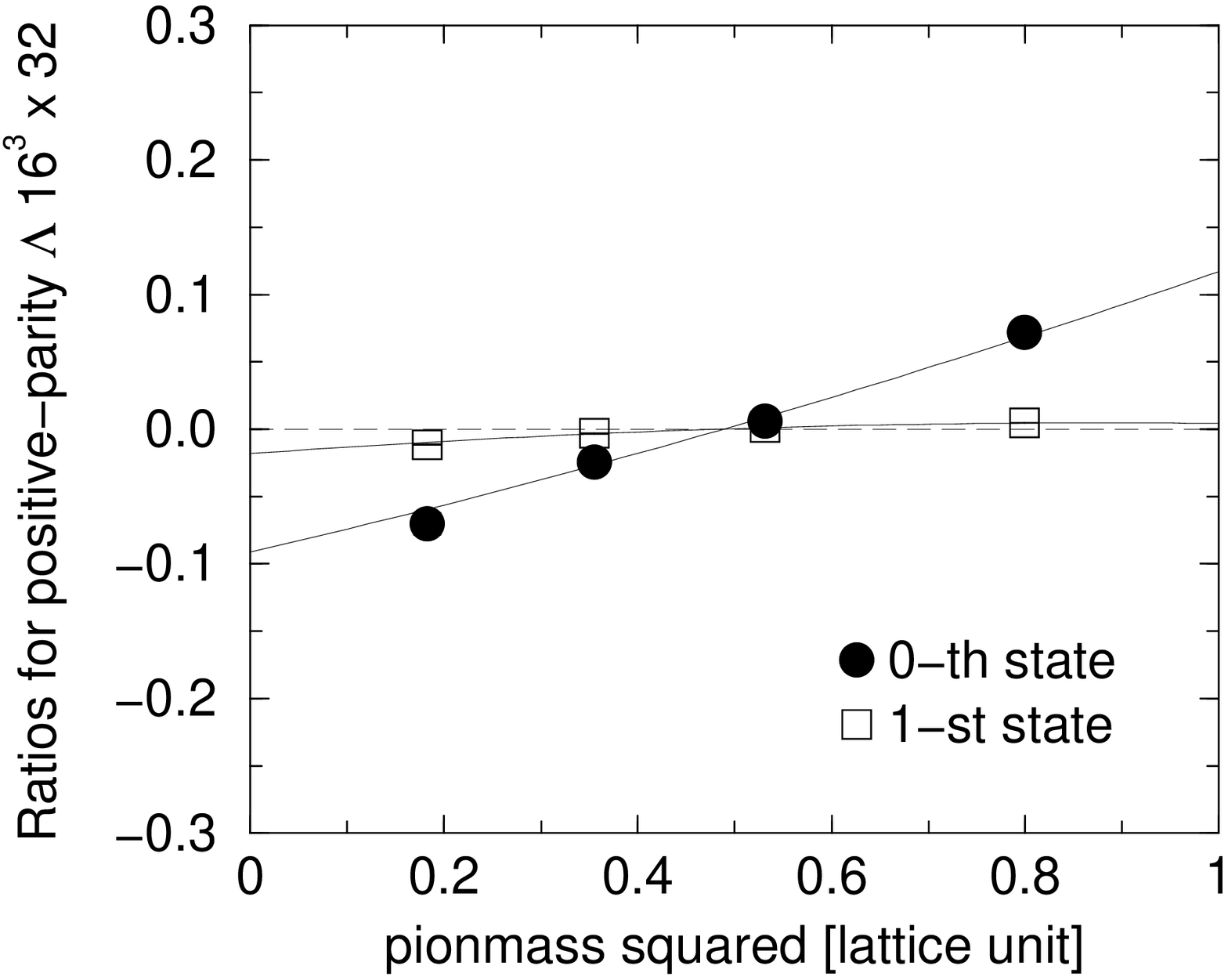}
\includegraphics[scale=0.35]{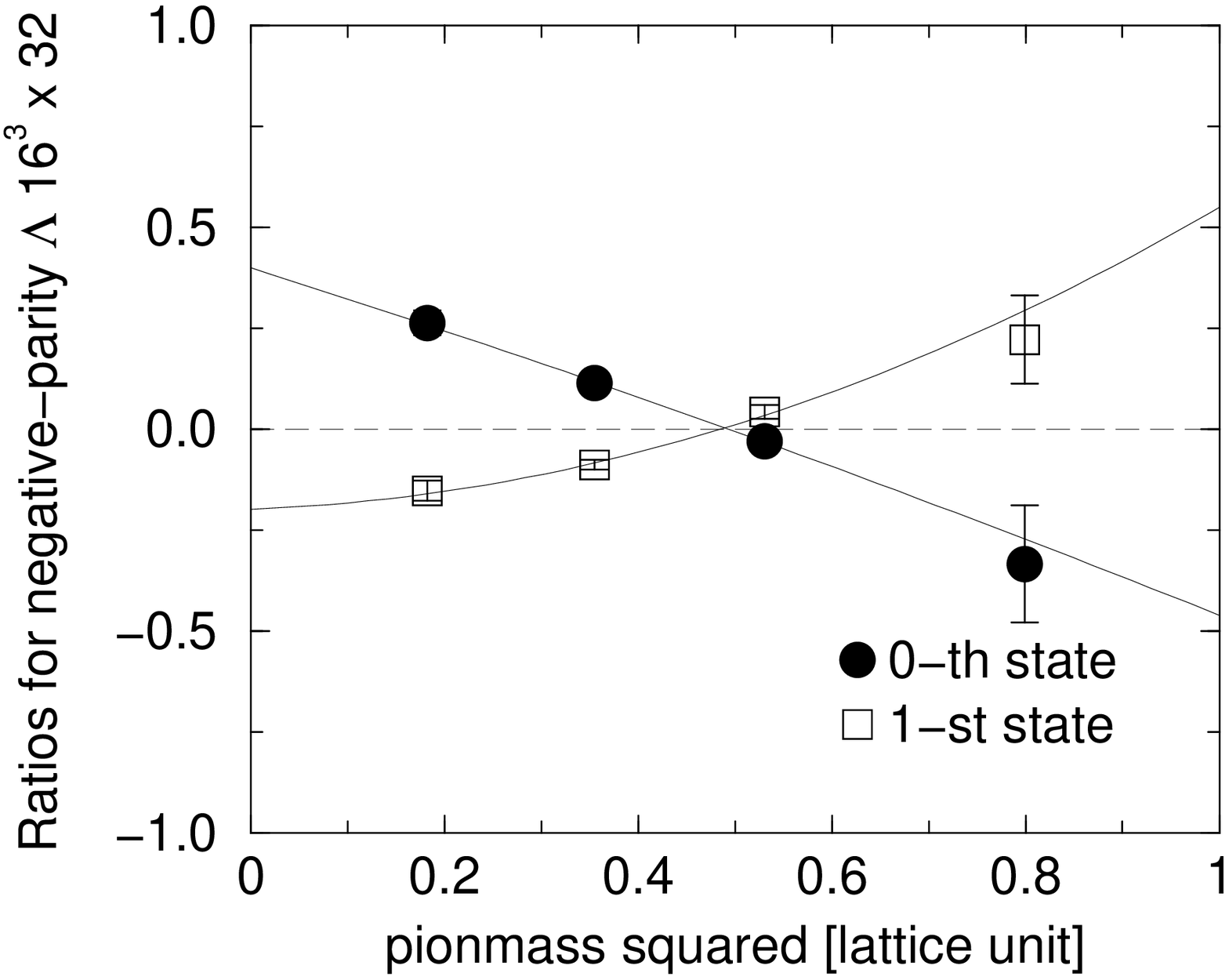}
\end{center}
\caption{\label{gpos}
$g_i^\pm$ are plotted as a function of $m_\pi^2$.
Solid lines denote quadratic functions used in chiral extrapolation.
({\bf left panel}): The ratio $g_i^+$ in the positive-parity channel.
({\bf right panel}): The ratio $g_i^-$ in the negative-parity channel.
}
\end{figure}

We show $g_0^\pm$ and $g_1^\pm$
as a function of pion-mass squared
for each lattice cutoff in Fig.~\ref{gpos}.
As is expected from the symmetry, such mixing coefficients
cross zero at the flavor symmetric limit
($\kappa_u=\kappa_d=\kappa_s$).
The data show smooth quark-mass dependence
toward the chiral limit in all the lattice-cut-off cases
and 
the dependences are almost lattice-cut-off independent.
We find that
the magnitude of operator mixing
gets larger for larger flavor-symmetry breaking.
It is interesting that
the 1st-excited state in positive-parity channel 
is dominated by the flavor-singlet component
showing almost no contaminations of octet components.

On the other hand, the mixings of the singlet and octet components
are generally weaker in the positive parity $\Lambda$'s.
It is natural because the splitting of the two states are large in the
SU(3)$_F$.
Nevertheless, it is interesting to see that the ground state, 
corresponding to $\Lambda(1115)$, contains significant (about 10\%) 
flavor singlet component in the chiral limit, which was not
expected from the simple quark model picture.

\section{Discussions}

Our results for negative-parity $\Lambda$ states indicate that 
there are two $1/2^-$ states nearly degenerate 
at around $1.6 -1.7$GeV,
while no state as low as $\Lambda (1405)$ is observed.
We have revealed the flavor structures of these states
from the lattice data for the first time.
It is found that 
the lowest negative-parity $\Lambda$ state is 
dominated by flavor-singlet component.
The second state, which is less than 100 MeV above the ground state, 
is mainly flavor octet.  
We thus find that the first two negative-parity $\Lambda$'s have different flavor
structures.

They, however, come significantly heavier than the experimentally observed
lowest-mass  state, $\Lambda(1405)$.
Considering that our simulation contains two-flavor dynamical quarks,
the failure of obtaining a light $\Lambda$ state could be attributed
either to
(1) strange-quark quenching,
(2) insufficient lattice volume or
(3) lack of chiral symmetry.
In fact, (1) seems to make the masses of octet baryons in positive
parity channel slightly ($\leq 10\%$) overestimated
in the present setups~\cite{AliKhan:2001tx}.
On the other hand, the deficiencies,(2) and (3), may cause the lowest
state not properly reproduced, supposing that the main component of
$\Lambda (1405)$ is a meson-baryon molecular state.
In order to check whether this conjecture is correct, simulations with
light dynamical quarks ($m_\pi \ll 500$ MeV)
and larger volume ($L \gg 2.5$ fm) will be required.

One may ask whether one or both of the obtained states 
in the present calculation 
correspond to the physical $\Lambda (1405)$.
One possible scenario is that
they correspond to the physical $\Lambda (1405)$ and $\Lambda (1670)$,
whose masses become larger due to the deficiencies discussed above.
While this scenario is simple and probable, 
we here consider two other possibilities.
First, suppose that these states correspond to the physical $\Lambda(1405)$, 
then the results may support its double pole structure 
proposed by the chiral unitary approach~\cite{Jido:2003cb}.
In our results, the lowest two states are almost degenerate at all the $\beta$'s (lattice spacing) and $\kappa$'s (quark masses).
Namely, the obtained two states are 
the signature of the double-pole resonance, but the
mass has not yet been reproduced because of the deficiency stated above.

The other possibility is that the obtained states do not correspond to the physical $\Lambda(1405)$, but
describe excited $\Lambda$ states.  In fact, the 2nd and 3rd negative-parity states lie at 1670 MeV
and 1800 MeV, both of which are three-star states in the Particle Data Group classification.
The present lattice data are consistent with these excited states. 
In the non-relativistic quark model approach, each of these states is classified as a flavor octet P-wave baryon.
We, however, have shown that the lower state is dominated by a flavor-singlet component.
So, the results here predict one flavor-singlet state and one flavor-octet state in the vicinity.

The flavor structures of the $\Lambda$ states
can be well clarified using the variational method.
The octet and the singlet components
are mixed when the flavor-SU(3) symmetry is broken:
The ground (1st-excited) state is dominated by singlet (octet) component, and 
the contamination by another representation
is at most 20\% (5\% when squared) in our present analysis.
From these findings, we expect that the flavor-SU(3) symmetry
is not largely broken.
A similar conclusion was 
also derived for the study of the meson-baryon coupling 
constants in lattice QCD~\cite{Erkol:2008yj}.
Because the SU(3) breaking effect seems not so large, 
the analyses without SU(3) mixings adopted so far
~\cite{Melnitchouk:2002eg,Nemoto:2003ft,Burch:2006cc,Ishii:2007ym}
make sense to some extent.
The mixings, however, get larger towards the chiral limit,
variational analyses could be essentially needed
when we adopt much lighter quarks.

\acknowledgments
All the numerical calculations were performed on NEC SX-8R at CMC, Osaka university and BlueGene/L at KEK. The unquenched gauge configurations employed in our analysis were all generated by CP-PACS collaboration~\cite{AliKhan:2001tx}. This work was supported in part by the 21st Century COE `Center for Diversity and University in Physics'', Kyoto University and Yukawa International Program for Quark-Hadron Sciences (YIPQS), by the Japanese Society for the Promotion of Science under contract number P-06327 and by KAKENHI (17070002, 19540275, 20028006 and 21740181).


\begin{thebibliography}{99}
\bibitem{Takahashi:2009bu}
  T.~T.~Takahashi and M.~Oka,
  arXiv:0910.0686 [hep-lat].

\bibitem{Sakurai:1960ju}
  J.~J.~Sakurai,
  Annals Phys.\  {\bf 11}, 1 (1960).

\bibitem{Dalitz:1967fp}
  R.~H.~Dalitz, T.~C.~Wong and G.~Rajasekaran,
  Phys.\ Rev.\  {\bf 153}, 1617 (1967).

\bibitem{Akaishi:2002bg}
  Y.~Akaishi and T.~Yamazaki,
  Phys.\ Rev.\  C {\bf 65}, 044005 (2002).

\bibitem{Yamazaki:2002uh}
  T.~Yamazaki and Y.~Akaishi,
  Phys.\ Lett.\  B {\bf 535} (2002) 70.

\bibitem{Akaishi:2005sn}
  Y.~Akaishi, A.~Dote and T.~Yamazaki,
  Phys.\ Lett.\  B {\bf 613}, 140 (2005)
  [arXiv:nucl-th/0501040].

\bibitem{Melnitchouk:2002eg}
  W.~Melnitchouk {\it et al.},
  Phys.\ Rev.\  D {\bf 67}, 114506 (2003)
  [arXiv:hep-lat/0202022].

\bibitem{Nemoto:2003ft}
  Y.~Nemoto, N.~Nakajima, H.~Matsufuru and H.~Suganuma,
  Phys.\ Rev.\  D {\bf 68}, 094505 (2003)
  [arXiv:hep-lat/0302013].

\bibitem{Burch:2006cc}
  T.~Burch, C.~Gattringer, L.~Y.~Glozman, C.~Hagen, D.~Hierl, C.~B.~Lang and A.~Schafer,
  Phys.\ Rev.\  D {\bf 74}, 014504 (2006)
  [arXiv:hep-lat/0604019].

\bibitem{Ishii:2007ym}
  N.~Ishii, T.~Doi, M.~Oka and H.~Suganuma,
  Prog.\ Theor.\ Phys.\ Suppl.\  {\bf 168}, 598 (2007)
  [arXiv:0707.0079 [hep-lat]].

\bibitem{AliKhan:2001tx}
  A.~Ali Khan {\it et al.}  [CP-PACS Collaboration],
  Phys.\ Rev.\  D {\bf 65}, 054505 (2002)
  [Erratum-ibid.\  D {\bf 67}, 059901 (2003)]

\bibitem{Sasaki:2001nf}
  S.~Sasaki, T.~Blum and S.~Ohta,
  Phys.\ Rev.\  D {\bf 65}, 074503 (2002)
  [arXiv:hep-lat/0102010].

\bibitem{Jido:2003cb}
  D.~Jido, J.~A.~Oller, E.~Oset, A.~Ramos and U.~G.~Meissner,
  Nucl.\ Phys.\  A {\bf 725} (2003) 181
  [arXiv:nucl-th/0303062].
  

\bibitem{Heller:1994rz}
  U.~M.~Heller, K.~M.~Bitar, R.~G.~Edwards and A.~D.~Kennedy,
  Phys.\ Lett.\  B {\bf 335}, 71 (1994)
  [arXiv:hep-lat/9401025].

\bibitem{Bolder:2000un}
  B.~Bolder {\it et al.},
  Phys.\ Rev.\  D {\bf 63} (2001) 074504
  [arXiv:hep-lat/0005018].

\bibitem{Erkol:2008yj}
  G.~Erkol, M.~Oka and T.~T.~Takahashi,
  Phys.\ Rev.\  D {\bf 79}, 074509 (2009)
  [arXiv:0805.3068 [hep-lat]].


\end{thebibliography}
\end{document}